\documentclass[nofootinbib,aps,prl,twocolumn,showpacs,superscriptaddress,amsmath]{revtex4-1}
\usepackage{dcolumn,bm}
\usepackage{natbib}
\usepackage{url}
\usepackage{hyperref}
\usepackage{graphicx}

\begin{document}

\title{Frequency Ratio of ${}^{199}$Hg and ${}^{87}$Sr Optical Lattice Clocks beyond the SI Limit}

\author{Kazuhiro Yamanaka}
\author{Noriaki Ohmae}

\affiliation{Quantum Metrology Laboratory, RIKEN, Wako-shi, Saitama 351-0198, Japan
}
\affiliation{Innovative Space-Time Project, ERATO, JST, Bunkyo-ku, Tokyo 113-8656, Japan
}
\affiliation{Department of Applied Physics, Graduate School of Engineering, The University of Tokyo, Bunkyo-ku, Tokyo 113-8656, Japan
}

\author{Ichiro Ushijima}
\author{Masao Takamoto}
\affiliation{Quantum Metrology Laboratory, RIKEN, Wako-shi, Saitama 351-0198, Japan
}
\affiliation{Innovative Space-Time Project, ERATO, JST, Bunkyo-ku, Tokyo 113-8656, Japan
}

\author{Hidetoshi Katori}
\affiliation{Quantum Metrology Laboratory, RIKEN, Wako-shi, Saitama 351-0198, Japan
}
\affiliation{Innovative Space-Time Project, ERATO, JST, Bunkyo-ku, Tokyo 113-8656, Japan
}
\affiliation{Department of Applied Physics, Graduate School of Engineering, The University of Tokyo, Bunkyo-ku, Tokyo 113-8656, Japan
}

\date{\today}

\begin{abstract}
We report on a  frequency ratio measurement of a ${}^{199}$Hg-based optical lattice clock referencing a ${}^{87}$Sr-based clock.
Evaluations of lattice light shift, including atomic-motion-dependent shift, enable us to achieve a total systematic uncertainty of $7.2 \times 10^{-17}$ for the Hg clock. 
The frequency ratio is measured to be $\nu_{\rm Hg}/\nu_{\rm Sr}=2.629\, 314\, 209\, 898\, 909\,  60(22)$ with a fractional uncertainty of $8.4 \times 10^{-17}$, which is smaller than the uncertainty of the realization of the SI second, i.e., the SI limit.
\end{abstract}


\maketitle

Rapid progress in optical lattice clocks~\cite{LeTargat2013,Hin13,Bloom2014,Fal14,Ushijima2014} makes them potential candidates for the future redefinition of the second and new tools for testing the fundamental laws of physics \cite{Uza03,Der14}.
The absolute frequencies of ${}^{87}$Sr and $^{171}$Yb-based optical lattice clocks have already been measured with the uncertainty of the realization of the SI second, allowing them to be adopted as the secondary representations of the second by Comit\`{e} International des Poids et Mesures (CIPM)~\cite{CIPM2013}.
Moreover, the systematic uncertainties of the ${}^{87}$Sr clock reach $10^{-18}$ level~\cite{Bloom2014,Ushijima2014}, which is competitive to ion-based clocks~\cite{Chou2010}. Such clocks can be used as new references to investigate  precise frequency measurements far beyond the SI second.

Hg is  another promising candidate for optical lattice clocks~\cite{Hachisu2008,McFerran2012a}, because its susceptibility to the blackbody radiation (BBR) is an order of magnitude smaller than that of Sr \cite{Bloom2014,Fal14,Ushijima2014} and Yb \cite{Beloy2014}.
Furthermore, a large nuclear charge of $Z = 80$ makes the Hg clock a sensitive probe for testing the constancy of the fine-structure constant~\cite{Angstmann2004}.
So far the absolute frequency  $\nu_{\rm Hg}$ of the ${}^1S_0-{}^3P_0$ clock transition of ${}^{199}$Hg has been reported with an uncertainty of $5.7 \times 10^{-15}$~\cite{McFerran2012a}, where the SI second is sufficient to describe the frequency.

In this Letter, we report on a frequency measurement of ${}^{199}$Hg by referencing to the ${}^{87}$Sr clock frequency $\nu_{\rm Sr}$~\cite{Ushijima2014}.
We have determined the frequency ratio to be $\nu_{\rm Hg}/\nu_{\rm Sr}=2.629\, 314\, 209\, 898\, 909\,  60(22)$ with a fractional uncertainty of $8.4 \times 10^{-17}$.
This ratio can be converted to  $\nu_{\text{Hg}} = 1\, 128\, 575\, 290\, 808\, 155.4(1.1) \, \textrm{Hz}$ via the recommended  frequency for the Sr transition $\nu_{\rm Sr}^{\rm CIPM} $\cite{CIPM2013} given by CIPM with uncertainty $1\times 10^{-15}$. 
To improve the systematic uncertainty of the Hg clock to $7.2 \times 10^{-17}$, we investigated the lattice light shift taking into account the multipolar polarizabilities~\cite{Katori2014}, which affect the light shift more seriously than in Sr and Yb clocks.

\begin{figure}[tb]
\includegraphics[width =\linewidth]{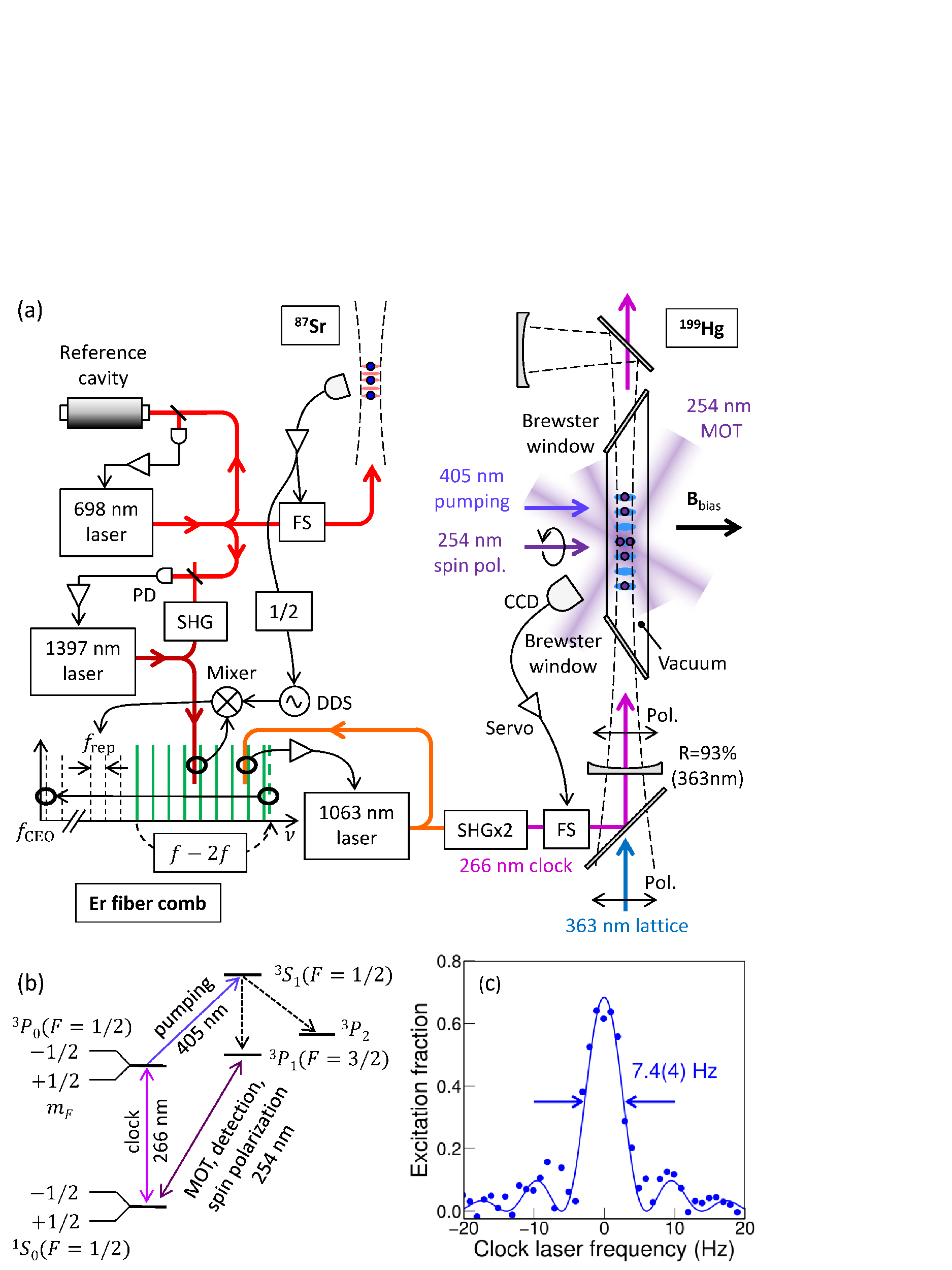}
\caption{
(a) Experimental setup.  Laser-cooled Hg atoms are loaded into an optical lattice at the magic wavelength of 363~nm.
The ${}^1S_0-{}^3P_0$ clock transition is interrogated by a clock laser at 266~nm, which is referenced to  the Sr clock laser at 698~nm via an optical frequency comb.
SHG, second harmonic generation; CCD, charge-coupled device; FS, frequency shifter; DDS, direct digital synthesizer.
(b) Relevant energy levels for $^{199}$Hg with a nuclear spin of $I=1/2$. (c) A clock spectrum of the ${}^1S_0-{}^3P_0$ transition of  $^{199}$Hg.
}
\label{fig:exp}
\end{figure}

Figure~\ref{fig:exp}(a) shows the experimental setup for the frequency ratio measurements, which consists of a ${}^{199}$Hg clock, a ${}^{87}$Sr clock~\cite{Ushijima2014}, and a frequency link between them.
The ${}^1S_0-{}^3P_0$  transition of $^{199}$Hg with a nuclear spin of $I=1/2$ [see Fig.~\ref{fig:exp}(b)] is used as the clock transition and is investigated by the following  experimental sequences with a cycle time of 1.5~s.
Hg atoms are laser-cooled by a vapor-cell type magneto-optical trapping (MOT) on the  ${}^1S_0-{}^3P_1$ transition at 254 nm with a natural linewidth $\mathit{\Gamma} = 1.3 \, \textrm{MHz}$~\cite{Hachisu2008}.
Initially, to collect atoms, we apply a MOT laser detuning $\delta \nu_{\rm MOT}\approx - 7 \mathit{\Gamma}$ and an intensity $\approx 10 \, {\rm mW/cm^2}$ per beam. 
After 860-ms-long atom loading time, we increase the gradient of quadrupole magnetic field from $1~{\rm mT/cm}$ to $3~{\rm mT/cm}$ to  compress the atom cloud. 
We finally tune the laser frequency to $\delta \nu_{\rm MOT}\approx- \mathit{\Gamma}$ and reduce the laser intensity to further cool the atoms to maximize the transfer efficiency into a lattice trap.

About 3 \% of atoms are loaded into a vertically-oriented one-dimensional (1D) optical lattice operated at $\lambda_{\rm L} \approx 362.6$~nm~\cite{McFerran2012a} with the maximum trap depth of $U_{\rm L}\approx 65\,E_{\rm R}$, where $E_{\rm R}/h = h/(2m\lambda_{\rm L}^2) = 7.6 \, {\rm kHz} $ is the lattice-photon recoil frequency, $h$ is the Planck constant, and $m$ is the mass of ${}^{199}$Hg atom.
We then temporarily decrease the trap depth down to $U_{\rm L}\approx35 \, E_{\rm R}$ to release the atoms trapped in high-lying axial vibrational states with $n\geq 3$.
This 1D lattice is formed inside a  buildup cavity with a power enhancement factor of $\approx 10$, which consists of two curved mirrors and a plane folding mirror  [see Fig.~\ref{fig:exp}(a)].
These cavity mirrors are placed outside the vacuum chamber to prevent vacuum-degradation of the mirror coating~\cite{McFerran2012a}, while the Brewster windows provide optical access to the vacuum and selective enhancement of the $p$-polarized light.
Atoms in the lattice are spin-polarized by exciting the ${}^1S_0-{}^3P_1$ transition with the circularly polarized light.
A  bias magnetic field ${\bf B}_{\rm bias} $ is applied during the spin polarization and clock excitation.

The ${}^1S_0-{}^3P_0$  transition is excited by a clock laser at 266~nm generated by two-stage frequency doubling of  a fiber laser at 1063~nm stabilized to an Er-doped fiber optical frequency comb by a linewidth transfer method~\cite{Inaba2013}.
The carrier envelope offset frequency $f_{\rm CEO}$ of the comb is stabilized by using a self-referencing $f-2f$ interferometer. 
The repetition rate $f_{\rm rep}$ of the comb is then stabilized by referencing a sub-harmonic (1397~nm) of a Sr clock laser  at 698~nm, which is prestabilized to a stable reference cavity with instability $\approx 5\times10^{-16}$ at $\tau=1$~s and stabilized to the Sr clock transition for $\tau >10$~s \cite{Ushijima2014}.

The clock laser is superimposed on the lattice laser with the same polarization.
Figure~\ref{fig:exp}(c) shows a clock spectrum with a  Fourier-limited Rabi linewidth of $\gamma_{\rm Hg} = 7.4(4) \, {\rm Hz}$, corresponding to a $Q$-factor of $\nu_{\rm Hg}/\gamma_{\rm Hg} = 1.5\times 10^{14}$, which is obtained for the clock interrogation time $\tau_{\rm i} = 120\,{\rm ms}$.
For the data presented below, we operate the Hg clock with $\tau_{\rm i} = 40-80 \, \textrm{ms}$ so that the frequency stabilization to the atomic transition becomes robust against the variations of experimental conditions for several hours.
The atom population $N_S$ in the $^1S_0$ state is determined from the  fluorescence  by operating the MOT for 20 ms.
We then optically pump the atoms in the ${}^3P_0$ state back to the ${}^1S_0$ state by exciting the ${}^3P_0-{}^3S_1$ transition at 405 nm to determine $N_P$ of atoms in the  ${}^3P_0$ state.
The excited atom fraction $\frac{N_P}{N_S+N_P}$ is used to stabilize the clock laser frequency.
We alternately interrogate the two $\pi$ transitions $m_F=\pm 1/2\rightarrow m_F=\pm 1/2$ to average out the 1st-order Zeeman shift and the vector light shift~\cite{Takamoto2006a}.
 
\begin{figure}[tb]
\includegraphics[width = \linewidth]{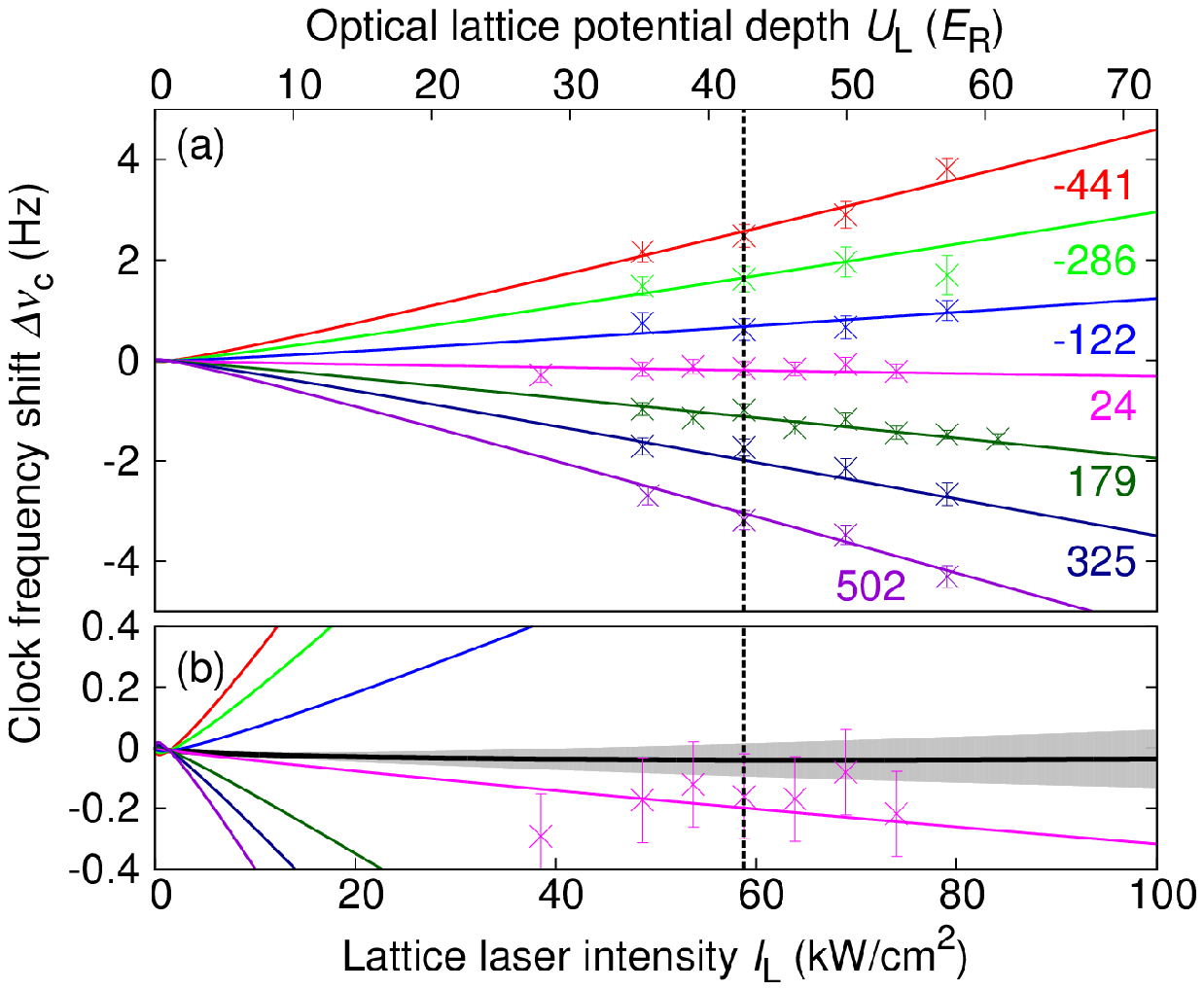} 
\caption{
Clock frequency shift $\Delta \nu_{\rm c} (I_{\rm L},\delta \nu_{\rm L})$ as a function of the  lattice laser intensity $I_{\rm L}$ (bottom axis). Corresponding lattice trap depth $U_{\rm L}$ is given in the top axis. 
 (a) Measured data (crosses) are fitted to Eq.~(\ref{eq:ls1}) shown by curves with corresponding colors.
 Numbers below the curves indicate the lattice laser frequency detuning $\delta \nu_{\rm L}$ (in MHz) from  $\nu_{\rm L}^{\rm E1} = 826\,855\,533\, {\rm MHz}$.
 (b) An enlarged view for the vertical axis.
For $\delta \nu_{\rm L} = -2\,{\rm MHz} $ used for the ratio measurements, 
the estimated clock frequency shift  and  uncertainty are shown by a black curve and a gray shaded region, respectively.
We operate the lattice at $I_{\rm L} \approx 59 \, {\rm kW/cm^2}$ or $U_{\rm L}\approx 43 E_{\rm R}$ as shown by a dashed line. 
 }
 \label{fig:ls}
 \end{figure}

The lattice light shift has been the primary source of the uncertainty of ${}^{199}$Hg clock~\cite{McFerran2012a}. 
To evaluate the lattice light shift $\Delta\nu_{\rm c}(I_{\rm L},  \nu_{\rm L})$, 
we measure the intensity-dependent  clock shift $\delta {\nu _{\rm{c}}}({I_{\rm{L}}},{I_0},{\nu _{\rm{L}}}) = \Delta {\nu _{\rm{c}}}({I_{\rm{L}}},{\nu _{\rm{L}}}) - \Delta {\nu _{\rm{c}}}({I_0},{\nu _{\rm{L}}})$ in  successive measurement cycles by varying the lattice laser intensity $I_{\rm L}$, while we keep $I_{0} = 89 \, {\rm kW/cm^2}$ constant.
Here the intensities $I_{\rm L}$ are given in terms of the peak intensity of a single traveling-wave laser that creates the lattice potential  depth of $U_{\rm L}$.
Figure~\ref{fig:ls}~(a) shows the data taken at 7 different lattice laser frequencies $\nu_{\rm L}$, which are stabilized to the optical frequency comb within $0.5 \, \textrm{MHz}$.
Each data point is measured with an uncertainty of $0.2\,{\rm Hz}$.

In the standing wave field, a spatial mismatch of the light shift  due to (i) the electric-dipole (E1) interaction and (ii) the electric-quadrupole (E2) and magnetic-dipole (M1) interactions introduces an atomic-motion-dependent light shift~\cite{Taichenachev2008}, which we refer to as the multipolar effect. 
In addition, the light shift due to the hyperpolarizability, coupled with the atomic motion in a lattice potential,  introduces non-linear light shift~\cite{Ovsiannikov2013}. 
Consequently, the total light shift shows intricate non-linear response to the light intensity  $I_{\rm L}$.

We estimate the leading hyperpolarizability shift $\Delta \beta I_{\rm L}^2$ to be  $ 2\times 10^{-17}$~\cite{Katori2014}  at our maximum lattice laser intensity of $I_{\rm L} = 89 \, {\rm kW/cm^2}$, which is smaller than our measurement uncertainties of $\sim 10^{-16}$. 
Therefore  we neglect the other hyperpolarizability effects than the leading term and approximate the clock shift as~\cite{Katori2014},
\begin{eqnarray}
&& h \Delta \nu_{\rm c} (I_{\rm L},\delta \nu_{\rm L},n) 
\approx
-\frac{\partial \Delta \alpha^{\rm E1}}{\partial \nu_{\rm L}}\delta \nu_{\rm L} I_{\rm L}(r) - \Delta \beta {I_{\rm L }(r)^2} \nonumber \\
&& +\left[ \frac{\partial \Delta \alpha^{\rm E1}}{\partial \nu_{\rm L}}\delta \nu_{\rm L} - \Delta \alpha^{\rm qm} \right] \left(n+\frac{1}{2}\right)  
\sqrt{ \frac{E_{\rm R}}{\alpha^{\rm E1}_0} I_{\rm L}(r) }. \label{eq:ls1}
\end{eqnarray}
Here, $n$ is the vibrational state for the axial motion, $\Delta  \alpha^{\rm E1}$ and $ \Delta \alpha^{\rm qm}$ are the differences of E1 and combined E2+M1 polarizabilities of the two clock states, and $\delta \nu_{\rm L}= \nu_{\rm L}-\nu_{\rm L}^{\rm E1}$ is the lattice laser detuning from the 
 ``E1-magic frequency'' $\nu_{\rm L}^{\rm E1}$ that equalizes E1 polarizabilities for the ground (g) and excited (e) states, i.e., $\alpha^{\rm E1}_0=\alpha^{\rm E1}_{\rm g (e)}$.
We note that the Gaussian intensity profile $I_{\rm L}(r) =  I_{\rm L} e^{-2r^2/w_0^2}$ of the lattice laser  confines atoms radially, where  $w_0$ is the beam radius and $r$ is the radial displacement.
For the typical lattice potential depth $U_{\rm L}\approx 43 E_{\rm R}$, 
the radial and the axial vibrational frequencies are $\nu_{\rm r} = \sqrt{4 U_{\rm L}/(\pi^2 m w_0^2)} \sim 100 \, {\rm Hz}$ and $\nu_{\rm a} = \sqrt{2 U_{\rm L}/(m \lambda_{\rm L}^2)} \sim 100 \, {\rm kHz}$. 
Since the typical kinetic energy  $\sim h \times100 \, {\rm kHz}$ of the lattice trapped atoms is about  $10^3$ times higher than the radial vibrational energy separation $h \nu_{\rm r}$, we treat the radial motion classically.
In addition, when calculating  an energy eigenvalue $E^{\rm vib}_{\rm g(e)} \left(I_{\rm L}(r),n\right)$ for the $n$-th axial vibrational state, the axial and radial motion may be decoupled as  the adiabaticity $\nu_{\rm a} \gg \nu_{\rm r}$ condition is satisfied.

To evaluate the light shift given in Eq.~(\ref{eq:ls1}), atomic motion in the lattice plays a crucial role, as the axial motion determines the averaged motional state $\left\langle n \right\rangle$ and the radial motion determines the effective lattice intensity $\left\langle I_{\rm L}(r) \right\rangle$ via the averaged atomic distribution ${\left\langle r^2 \right\rangle}$ in the Gaussian intensity profile.
The axial vibrational population in the $n=0$ state is measured to be $0.7(1)$ by the difference of the total area of the red and blue motional sideband spectra \cite{Lei03}.
As atoms in $n\geq3$ are removed in the state preparation,  the average occupation is estimated to be $\left\langle n \right\rangle  = 0.4(2)$ by assuming the atoms are Boltzmann-distributed among $n=0,1,2$ vibrational states.
The radial atomic distribution  can be inferred from the inhomogeneously broadened sideband lineshapes~\cite{Blatt2009}, as the sideband frequency becomes smaller as ${\left\langle r^2 \right\rangle}$ increases due to the  Gaussian intensity distribution $I_{\rm L}e^{-2r^2/w_0^2}$ of the lattice laser.
The blue-sideband frequency $\nu_{\rm b} (n,r) =  [E^{\rm vib}_{\rm e} (I_{\rm L}(r),n+1) -E^{\rm vib}_{\rm g} (I_{\rm L}(r),n)] /h$ is given by,
\begin{eqnarray}
\nu_{\rm b}(n, r) 
&\approx&
 \nu_{\rm a}e^{-r^2/w_0^2} - E_{\rm R}(n + 1)/h.
\end{eqnarray}
Using the sideband spectrum, we determine $\langle e^{-2r^2/w_0^2}\rangle= 0.8(1)$.

We experimentally measure  $\alpha^{\rm E1}_0/h=5.5 (8) \, {\rm kHz/(kW/cm^2)}$ from the axial motional sideband frequency, where the uncertainty is given by the measurement uncertainty of the lattice laser intensity. 
We use calculated value $\Delta \alpha^{\rm qm}/h = 8.25 \,{\rm mHz/(kW/cm^2)}$ and $\Delta \beta /h = -2.5 \, {\rm \mu Hz/(kW/cm^2)^2}$~\cite{Katori2014}.
The entire data are then fitted to Eq.~(\ref{eq:ls1}) (see Fig.~\ref{fig:ls}), where we employ a multiple regression analysis taking $I_{\rm L}$ and $\nu_{\rm L}$ as explanatory variables and $\Delta\nu_{\rm c}$ as a response variable.
This determines the E1-magic frequency of $\nu_{\rm L}^{\rm E1} = 826\,855\,533(9)\, {\rm MHz}$ and $\frac{1}{h}\frac{\partial \Delta \alpha^{\rm E1}}{\partial \nu_{\rm L}} = 1.5(4) \times 10^{-10} \, {\rm /(kW/cm}^2)$.

For the measurement of Hg clock frequency, we take the lattice frequency to be $\nu_{\rm L} = \nu_{\rm L}^{\rm E1} - 2 \, {\rm MHz} = 826\,855\,531 \, {\rm MHz}$ so that the light shift $\Delta \nu_{\rm c}(I_{\rm L},\nu_{\rm L})$ becomes  insensitive, i.e., ${\left. {{\textstyle{{\partial \Delta {\nu _{\rm{c}}}({I_{\rm{L}}},{\nu _{\rm{L}}})} \over {\partial {I_{\rm{L}}}}}}} \right|_{{I_{\rm{L}}} = {I_{{\rm{op}}}}}} = 0$,  to the variation of the lattice intensity around $I_{\rm op}= 59(10) \, {\rm kW/cm^2}$.
The relevant light shift (black  curve) and its uncertainty (gray shaded area) are shown in Fig.~\ref{fig:ls}(b). 
The lattice light shift is estimated to be $\Delta \nu_{\rm c} = -0.04(7)\, {\rm Hz}$  corresponding to the fractional frequency shift of $-4(6) \times 10^{-17}$. 
In order to further investigate the lattice light shift, experimental determinations of $\Delta \beta$ and $\Delta \alpha^{\rm qm}$ are crucial. 

\begin{table}[tb]
\caption{\label{tab:budget}Corrections and uncertainties for $\nu_{\rm Hg}/\nu_{\rm Sr}$ measurement.
All numbers are shown in a fractional unit of $10^{-17}$.}
\begin{ruledtabular}
\begin{tabular}{ldddd}
 Effect & \multicolumn{1}{c}{Correction} & \multicolumn{1}{c}{Uncertainty} \\
\hline 
Lattice light  & 3.6	&	6.1 \\ 
BBR & 16.1	& 3.3 \\ 
Atom density & 1.6 &  1.6  \\ 
2nd-order Zeeman & 6.1 	& 0.9\\ 
Clock light & 0	& < 0.1  \\ 
AOM chirp & 0 & < 1  \\ 
Servo error & -0.4 &	0.3 \\ 
\hline
\textbf{Hg clock systematic total} & 26.9 & 7.2 \\ 
\hline 
Sr clock systematic	& -17.0 & 0.7	\\
Gravitational redshift	& -0.5	& 0.1 \\
1st-order Doppler	& 0	& < 2		\\
Statistic 	& -		& 3.7	\\
\hline
\textbf{Total}	& 9.4		& 8.4
\end{tabular}
\end{ruledtabular}
\end{table}

As listed in Table~\ref{tab:budget}, effects other than the lattice light shifts give relatively minor contributions to the systematic uncertainty of the Hg clock.
The BBR shift is estimated to be $-1.6(3) \times 10^{-16}$ for the ambient temperature of 297(3)~K, where the uncertainty mainly comes from the theoretical polarizabilities with 10\% uncertainty~\cite{Hachisu2008}.
The atom density shift is measured by varying the number of atoms by $50\%$.
A linear fit to the data points infer the collisional shift of $-2(2) \times 10^{-17}$ for  $N\approx 1000$ atoms in the lattice,  corresponding to the atom density of $\approx 1 \times 10^{10} \, {\rm cm^{-3}}$ or an average single-lattice-site occupation of 1.2 atoms.
The second-order Zeeman shift is investigated by varying ${{\bf{B}}_{{\rm{bias}}}}$ in successive measurement cycles, where the magnetic field is measured through the  first-order Zeeman shift  and the g-factor difference in  the clock transition~\cite{Lahaye1975}.
Fitting to $\Delta \nu_{\rm c}  = -\beta_Z {|{\bf{ B}}_{\rm bias}|}^2$ yields $\beta_Z = 1.6(2) \, {\rm Hz/(mT)^2}$, giving a shift of $-6(1) \times 10^{-17}$ at our operating condition of $|{\bf{ B}}_{\rm bias}| = 0.21 \, {\rm mT} $.
The light shift induced by the clock laser is estimated to be $< 1 \, {\rm mHz}$~\cite{Wexler1980}. 
The servo error is estimated to be smaller than the statistical uncertainty of the measurement for an averaging time $\tau > 100 \,{\rm s}$ by analyzing the error signal of the frequency stabilization to the clock transition.

Sources of uncertainties for the frequency ratio measurement are also listed in Table~\ref{tab:budget}.
The Sr clock achieves the systematic uncertainty of $7\times10^{-18}$~\cite{Ushijima2014}.
The gravitational redshift is estimated from the height difference of $\Delta h = 5(1) \, {\rm cm}$ between the two clocks.
The first-order Doppler shift between the Hg and the Sr clocks is estimated to be less than $2\times 10^{-17}$, which is due to the temperature drift in the uncompensated optical path of about $ 5 \, {\rm m}$.

\begin{figure}[tb]
\includegraphics[width = \linewidth]{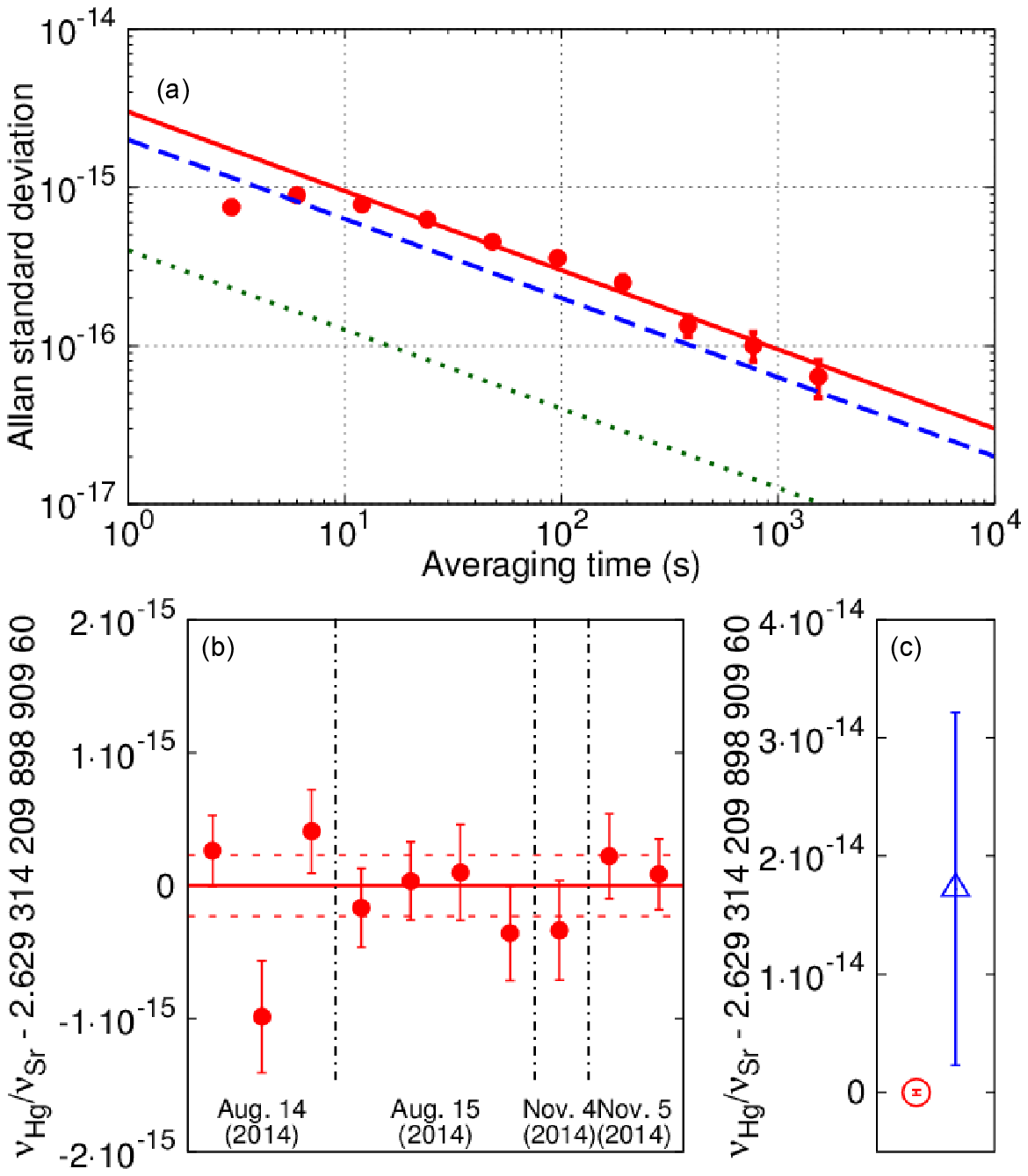} 
\caption{(a) The Allan standard deviation for  the ratio measurement $\nu_{\rm Hg} / \nu_{\rm Sr}$ (circles).
Solid line shows a $ \sigma_y(\tau) = 
3\times 10^{-15} / \sqrt{\tau/{\rm s}}$ trend.
 Dashed and dotted lines show the Dick-effect-limited and QPN-limited instabilities.
(b) The ratio  measurements  carried out with a 3-month-interval, where the solid and dashed lines show the weighted mean and the total uncertainty of our measurements.
(c) Summary of the ratio measurements. 
Our result (red circle) is displayed with that reported by LNE-SYRTE group (blue triangle) $\nu_{\rm Hg}^{\rm SYRTE}/\nu_{\rm Sr}^{\rm SYRTE}$. 
Error bars indicate $1\sigma$ uncertainties.
}
\label{fig:ratio}
\end{figure}

Figure~\ref{fig:ratio}(a) displays the Allan standard deviation for the ratio measurement $\nu_{\rm Hg} / \nu_{\rm Sr}$, which  shows a $ \sigma_y(\tau) = 
3\times 10^{-15} / \sqrt{\tau/{\rm s}}$ trend for an averaging time of $\tau \geq$ 10~s, mainly responsible to the Dick-effect-limited instability~\cite{Santarelli1998} of $ \sigma_y^{\rm Dick}(\tau)\approx 2 \times 10^{-15}/\sqrt{\tau / {\rm s}}$. 
The quantum projection noise (QPN)~\cite{Itano1993} only contributes to the instability of $ \sigma_y^{\rm QPN}(\tau)\approx 
4 \times 10^{-16}/\sqrt{\tau / {\rm s}}$.
Application of  a clock laser with smaller instability or synchronous operation of two clocks to reject frequency noise of the clock laser~\cite{Takamoto2011} will allow approaching the QPN-limited instability.

Figure~\ref{fig:ratio}(b) summarizes the frequency ratio measurement $\nu_{\rm Hg}/\nu_{\rm Sr}$.
For the evaluation of the statistical uncertainty of $\nu_{\rm Hg}/\nu_{\rm Sr}$, we construct a histogram of all the data with the bin size of $5\times 10^{-16}$ for a fractional frequency. 
The histogram is fitted to the normal distribution function, where the reduced chi-squared yields $\sqrt{\chi_{\rm red}^2} = 4.0$.
We thus conservatively evaluate the statistical uncertainty to be $3.7 \times 10^{-17}$ by inflating the standard uncertainty of the mean by $\sqrt{\chi_{\rm red}^2}$. 
A weighted mean of the result is $\nu_{\rm Hg}/\nu_{\rm Sr} = 2.629\, 314\, 209\, 898\, 909\,  60(22) $, where the fractional uncertainty of $8.4 \times 10^{-17}$ is essentially given by the systematic uncertainty of the Hg clock (see Table~\ref{tab:budget}).
The ratio $\nu_{\rm Hg}^{\rm SYRTE}/\nu_{\rm Sr}^{\rm SYRTE} $, both of which are measured by referencing Cs primary standards at LNE-SYRTE~\cite{McFerran2012a,LeTargat2013}, is shown by a triangle in Fig.~\ref{fig:ratio}(c).
This result is consistent with our measurements to about $1\sigma$ uncertainty.
Applying the recommended value $\nu^{\rm CIPM}_{\rm Sr}$ of ${}^{87}$Sr as a secondary representation of the second~\cite{CIPM2013}, the absolute frequency of ${}^{199}$Hg is given by $\frac{\nu_{\rm Hg}}{\nu_{\rm Sr}}\cdot \nu^{\rm CIPM}_{\rm Sr} = 1\, 128\, 575\, 290\, 808\, 155.4(1.1) \, \textrm{Hz}$ in the unit of  SI second, where an uncertainty of $1\times 10^{-15}$ is  given by that of $\nu^{\rm CIPM}_{\rm Sr} $.

In summary, we have investigated an optical lattice clock based on ${}^{199}$Hg and achieved a total systematic uncertainty of $7.2 \times 10^{-17}$. 
We determine the magic frequency by including the multipolar effect that seriously affects the uncertainty budget at low $10^{-16}$. 
Similar strategy will be applied for  Sr and  Yb-based clocks, where the  multipolar and hyperpolarizability effects become relevant for the systematic  uncertainty at the $10^{-17}$ level~\cite{Katori2014}.
We have determined the frequency ratio between optical lattice clocks $\nu_{\rm Hg}/\nu_{\rm Sr}$ with an uncertainty of $8.4 \times 10^{-17}$.
Accurate determinations of such ratios allow the investigation of  the constancy of the fine-structure constant $\alpha$. Taking the Sr clock as an $\alpha$-insensitive anchor, the fractional change $\Delta \nu$ of the Hg clock reveals $\Delta \alpha/\alpha  = 0.8 \Delta \nu/\nu_{\rm Hg}$~\cite{Angstmann2004}, which will be competitive to the previous  constraints~\cite{Ros08} if $\Delta \nu/\nu_{\rm Hg}\approx 10^{-17}$ is investigated over a year.

\begin{acknowledgments}
We thank F.-L. Hong, H. Inaba, Y. Kaneda, and P. Thoumany for laser developments, and T. Akatsuka, M. Das, N. Nemitz, T. Pruttivarasin, T. Takano, and A. Yamaguchi for useful comments and discussions.
This work was partly supported by the FIRST Program of the JSPS and by the Photon Frontier Network Program of the MEXT, Japan.
K. Y. acknowledges financial support from Grant-in-Aid for JSPS Fellows and ALPS.
\end{acknowledgments}

\bibliographystyle{apsrev4-1}
\bibliography{HgSrpaper}

\end{document}